\documentclass{IEEEtran}
\usepackage{graphicx}
\usepackage{epsfig}
\usepackage{amsmath}
\usepackage{cite}
\usepackage{booktabs}

\usepackage{amsmath,amssymb,amsfonts}
\usepackage{algorithmic}
\usepackage{graphicx}
\usepackage{textcomp}

\usepackage[colorlinks,linkcolor=blue,citecolor=blue]{hyperref}   
\newcommand{\ba}{{\ensuremath{\boldsymbol{a}}}}

\newcommand{\bs}{{\ensuremath{\boldsymbol{s}}}}

\newcommand{\bD}{{\ensuremath{\boldsymbol{D}}}}
\newcommand{\bA}{{\ensuremath{\boldsymbol{A}}}}
\newcommand{\bB}{{\ensuremath{\boldsymbol{B}}}}

\newcommand{\bS}{{\ensuremath{\boldsymbol{S}}}}
\newcommand{\bU}{{\ensuremath{\boldsymbol{U}}}}
\newcommand{\bV}{{\ensuremath{\boldsymbol{V}}}}
\newcommand{\bSigma}{{\ensuremath{\boldsymbol{\Sigma}}}}

\title{\LARGE\textbf{ A Singular-value-based Marker for the Detection of Atrial Fibrillation Using High-resolution Electrograms and Multi-lead ECG}}

\author{Hanie Moghaddasi\thanks{Corresponding author: h.moghaddasi@tudelft.nl}$^1$, Richard C. Hendriks$^1$, Borb\'{a}la Hunyadi$^1$, Paul Knops$^2$, Mathijs S van Schie$^2$,\\ Natasja M.S. de Groot{$^1$,$^2$}, and Alle-Jan van der Veen$^1$\\

$^1$Delft University of Technology, Delft, The Netherlands \\
$^2$Erasmus University Medical Center, Rotterdam, The Netherlands}

\begin{document}
\maketitle

\begin{abstract}
  \textit{Objective:} The severity of atrial fibrillation (AF) can be assessed from intra-operative epicardial measurements (high-resolution electrograms), using metrics such as conduction block (CB) and continuous conduction delay and block (cCDCB).
These features capture differences in conduction velocity and wavefront propagation. However, they do not clearly differentiate patients with various degrees of AF while they are in sinus rhythm, and complementary features are needed.  In this work, we focus on the morphology of the action potentials, and derive features to detect variations in the atrial potential waveforms. \textit{Methods:} We show that the spatial variation of atrial potential morphology during a single beat may be described by changes in the singular values of the epicardial measurement matrix. The method is non-parametric and requires
little preprocessing. A corresponding singular value map points at areas subject to fractionation and block. Further, we developed an experiment where we simultaneously measure electrograms (EGMs) and a multi-lead ECG. \textit{Results:} The captured data showed that the normalized singular values of the heartbeats
during AF are higher than during SR, and that this difference is
more pronounced for the (non-invasive) ECG data than for the EGM data, if the
electrodes are positioned at favorable locations. 
\textit{Conclusion:} Overall, the singular value-based features are a useful indicator to detect and evaluate  AF. \textit{Significance:} The proposed method might be beneficial for identifying electropathological regions in the tissue without estimating the local activation time.
\end{abstract}

\section{Introduction}
Atrial fibrillation (AF) is the most prevalent and persistent cardiac tachyarrhythmia. In the electrocardiogram (ECG), AF is characterized by fibrillatory waves and irregular RR intervals. Various mechanisms have been proposed to underly AF, including multiple wavelets, rotors, re-entrant activity, endo-epicardial breakthrough waves, and ectopic foci \cite{guillem2016presence}, but the precise pathological mechanisms causing AF in the individual AF patient are yet unknown.  

High-resolution electrograms (EGMs) are used to understand the electropathological process of AF in more detail. From such measurements, the electrical propagation in the heart and the conduction velocity are assessed by the local activation time (LAT) and derived features, such as conduction block (CB) and continuous conduction delay and block (cCDCB). This analysis is influenced by the accuracy of the LAT estimation. Furthermore, these features assessed during sinus rhythm do not sufficiently differentiate between the various stages of AF development \cite{van2021conduction, lanters2017spatial}. 

In some cases, the electropathology of atrial tissue can also be linked to the morphology of the observed signals. The R/S ratio of single potentials (SPs) has been shown to be useful for assessing the severity of conduction inhomogeneity \cite{ye2021signal,van2020classification}. This is complementary to the LAT analysis: a wavefront\footnote{To compare the various wavefronts visually, see Fig. \ref{LAT_sing_sim}} could propagate normally, even while the underlying APs are abnormal. In our previous work, we have demonstrated that the development stages of AF might manifest themselves as variations in atrial potential waveforms \cite{moghaddasi2022novel}. 

In this paper, we go one step deeper and study a measurement data matrix formed from an array of unipolar EGMs. This matrix is preprocessed such that it becomes insensitive to differences in LAT but remains sensitive to spatial differences in AP morphology. We then look at the singular values of this matrix. A simple cardiac signal model demonstrates that if all cells beneath the electrodes generate the same action potential (AP), and the propagation follows a flat wavefront, then the data matrix has rank 1: only a single singular value is nonzero. Furthermore, with a slightly more elaborate signal model, we will demonstrate that, in case of more complex underlying physiology - such as differences in AP morphology or abnormal wavefront propagation - the data matrix has a higher rank, which can be detected by an increase in the second singular value. 

Thus, our hypothesis is that the normalized second singular value is a
useful feature to detect and classify degrees of AF. This is tested on
two types of clinical data.
First, we study this feature on measured intraoperative unipolar EGMs obtained from patients with induced AF. Comparing SR and AF data, we found that the normalized singular values are significantly higher during AF than during SR, and this allows to  discriminate between SR and AF.

Next, we also study singular values for data collected using a (non-invasive) multi-lead ECG. In particular, we have designed a sub-vest to monitor the body surface potentials (BSPs) at 15 leads simultaneously combined with EGM mapping at specific epicardial locations during minimally invasive surgery. This vest was designed by the authors as a standard 12-lead ECG cannot be simultaneously acquired during open-heart surgery. The acquired data allows to study the singular values of the low-resolution BSP data for exactly the same heartbeats as the high-resolution EGMs. The results show that, for specific placements of the electrodes, the BSP shows even more significant changes in singular values than the EGM, presumably because the latter only measures a small area of the atrial surface.

For EGM data acquired on a sufficiently large rectangular grid, the normalized second singular value can also be computed from overlapping submatrices of $3\times 3$. Looking at the submatrices helps us to locate the abnormal electropathological regions in the tissue. This allows to construct a novel map that has complementary information to the traditional activation map. The results show that this map highlights areas of double potentials, simultaneous presence of multiple AP morphologies, and block. These are often associated with AF. We propose this map as a useful tool, complementary to the use of activation maps.

A related eigenvalue analysis (and corresponding map) was proposed by Riccio e.a.\ \cite{riccio2022atrial}. As a preprocessing step, their method requires to time-align the time-domain electrogram traces. The estimation of the local activation times is an additional step, which also based on an underlying model where for each trace a single activation time can be estimated. This is problematic at electrodes that see a double potential. In contrast, our proposed method requires only little preprocessing.

The rest of this paper is structured as follows. In Section \ref{sec2}, we introduce our method, including notation, action potential, and electrogram model, and analyze the singular values in relation to various scenarios, such as one or more signal morphologies, and one or more wavefronts. In Section \ref{sec3} we demonstrate the proposed approach on simulated data and two types of clinical data. In Section \ref{sec4}, we discuss the results. Finally, conclusions are phrased in Section \ref{sec5}.

\section{Methods and algorithms}\label{sec2}
\subsection{Notation} \label{not}
In this paper, scalars are denoted by normal lowercase letters, vectors by bold lowercase letters, and matrices by bold uppercase letters.  
For matrices, $(|\cdot|)$ is the element-wise absolute value, and $(.^H)$ denotes the Hermitian operator.

\subsection{Action potential and electrogram model}
An action potential is generated by a sequence of voltage changes across the membrane of a cell. Various mathematical models have been proposed to describe the AP in atrial myocytes and pacemaker cells. In particular, the total ionic current in human atrial myocytes can be computed from the Courtemanche model \cite{courtemanche1998ionic}, while for pacemaker cells, which have funny currents, the ionic current is governed by the Fabbri et al.\ model \cite{fabbri2017computational}.\footnote{The Courtemanche model was used to produce the simulated data for our analysis. By varying parameters, this model can generate a variety of AP morphologies. More elaborate computer models of electrograms have been developed in \cite{virag2002study,jacquemet2006analysis,abdi2019compact}, and these could be used to improve our analysis.}
Next, a reaction-diffusion equation models the AP propagation in a 2D tissue, described by the interaction of three currents: the transmembrane current, the stimulus current, and the ionic current. 
In models with uniform parameters, the resulting APs are the same for all cells, and a simple data model to describe this is
\begin{equation}\label{ym}
        d_c(t) = a_c\, s(t-\tau_c)\,,\quad c = 1, \cdots, N_c\,, 
\end{equation}
where $d_c(t)$ is the AP (voltage) for the $c$th cell, $a_c$ is a positive real amplitude, $s(t)$ is the reference AP, and $\tau_c$ is the time delay between a reference cell and the $c$th cell. $N_c$ denotes the total number of cells in the 2D tissue model. Here, ``cell'' does not refer to a physical cell, but rather a space-discretized grid point representing a collection of physical cells. The delays $\tau_c$ are the LATs. These are related to each other via wavefronts, which follow from the selected diffusion-reaction model, the underlying conductivity tensors, and an initial stimulus scenario \cite {virag2002study}.

Moving one level up, the electrogram as measured on the epicardium is modeled by a collection of $M$ electrodes assumed to be placed at a constant height above the 2D tissue. Each electrode measures a weighted sum of the action potentials from all cells on the 2D tissue. The $m$th electrode signal (voltage) $d_m(t)$ at location $(x_m,y_m)$ with a constant height $z_0$ above the 2D tissue then can be modeled by a space-discretized equation \cite{abdi2019compact} as
\begin{equation}\label{egm}
\begin{aligned}
   &d_m(t)= \sum_{c=1}^{N_c} a_{m,c}\, d_c(t)
   \,,\qquad m = 1, \cdots, M
   \\
   & a_{m,c} = \frac{a}{\sqrt{(x_c-x_m)^2+(y_c-y_m)^2+z^2_0}} \,.
\end{aligned}
\end{equation}
Here, the weight $a_{m,c}$ describes the instantaneous gain from cell $c$ to electrode $m$, using an inverse relation to distance, and $a$ is a constant scale parameter (electrode gain). The electrode size in this model formulation is very small and is considered as a point electrode.

\subsection{Data stacking and processing}
Returning to the cell model (\ref{ym}), we first apply a Fourier transform: let 
\begin{equation}\label{ym_ff}
       \Tilde{d}_c(\omega)=\int_{-\infty}^{+\infty}d_c(t)e^{-j\omega t}dt
\end{equation}
where $(\Tilde{.})$ denotes the frequency domain.
We take $N$ samples in frequency domain:\footnote{In practice, we would sample in time domain and use the FFT.} 
$\omega \in \{\omega_1,\omega_2,\cdots, \omega_N\} $. The $N_c\times N$ complex samples are stacked into a matrix $\bD$:
\[
  \bD = [\Tilde{d}_c(\omega_n)]_{c,n}
\;\in\; \mathbb{C} ^{N_c\times N} \,.
\]
For electrode signals $d_m(t)$, we can do a similar processing, resulting in a matrix that we also denote by $\bD$, but that now will have size $M\times N$.

As motivated later, we drop the phase by taking the element-wise absolute value of $\bD$:
\[
   \bB = | \bD| \,.
\]
Finally, we compute the singular value decomposition (SVD) of $\bB$ as
\begin{equation}   \label{svd}
   \bB =\bU \bSigma \bV^H \,,
\end{equation}
where $\bU$ and $\bV$ are unitary matrices containing the left and right singular vectors, and $\bSigma$ is a diagonal matrix containing the singular values $\{\sigma_1,\cdots,\sigma_N\}$, sorted in descending order.

The singular values are indicative of the numerical rank of the matrix, and they give important information on the complexity of the matrix. Next, we analyze these singular values for several cases of interest.

\subsection{Singular value analysis}
\subsubsection{Cell level, single AP morphology}
   We start at the cell level and assume, as in (\ref{ym}), that under healthy conditions all cells generate APs with the same morphology. In the frequency domain, (\ref{ym}) gives 
\begin{equation}\label{ym_f}
        \Tilde{d}_c(\omega) = a_c\, e^{-j\omega\tau_c} \Tilde{s}(\omega) \,.
\end{equation}
   After discarding the phase by taking the absolute value, and stacking the magnitude spectra into the matrix $\bB$, we observe that
\begin{equation}\label{y_sr}
        \bB = \ba \bs^T \,,
\end{equation}
   where $\ba = [a_1,\cdots,a_{N_c}]^T$ and 
   $\bs^T = [|\Tilde{s}(\omega_1)|,\cdots,|\Tilde{s}(\omega_N)|]$.
   This model shows that $\bB$ has rank 1: only one singular value is non-zero. 
   This important property is achieved by discarding the phase (which contains the effect of the LATs), and the assumption that all cells have the same AP morphology. 
   Unfortunately, some information on the morphology is lost, since we also drop the phase of $\Tilde{s}(\omega)$.

    Since the LATs do not play a role after taking the absolute value, it does not matter whether the AP model describes an SR scenario ($\tau_c$ organized in a single wavefront) or an AF scenario ($\tau_c$ organized in multiple wavefronts, or highly unstructured).

\subsubsection{Cell level, two different AP morphologies}
    As a second case, we consider a scenario where cells take one out of two morphologies, $s_1(t)$ or $s_2(t)$. 

    The data model for $\bB$ results in
\begin{equation}   \label{B2}
   \bB = \bA\bS = \ba_1 \bs_1^T + \ba_2 \bs_2^T
\end{equation}
    where $\bA = [\ba_1\;\; \ba_2]$ and $\bS = [\bs_1\;\;\bs_2]^T$.
    Entries of $\ba_1$ are zero if the corresponding cell is of the second type, and likewise, entries of $\ba_2$ are zero if a cell is of the first type. Thus, the columns of $\bA$ are complementary and trivially orthogonal. Since by assumption $\bs_1 \neq \bs_2$, the matrix $\bB$ has rank 2, and only two singular values are nonzero.

    These two singular values are determined by two parameters:
    \begin{enumerate}
        \item The difference between $\bs_1$ and $\bs_2$, as expressed by their cross-correlation. If the difference is small, then the second singular value will be small.
        \item The number of cells in group 1 versus the number of cells in group 2: this determines the ratio $\|\ba_1\|/\|\ba_2\|$. If the cells are predominantly in one group, then the second singular value will be small.
    \end{enumerate}

\begin{figure*}
    \centering
    \includegraphics[width=0.9\textwidth]{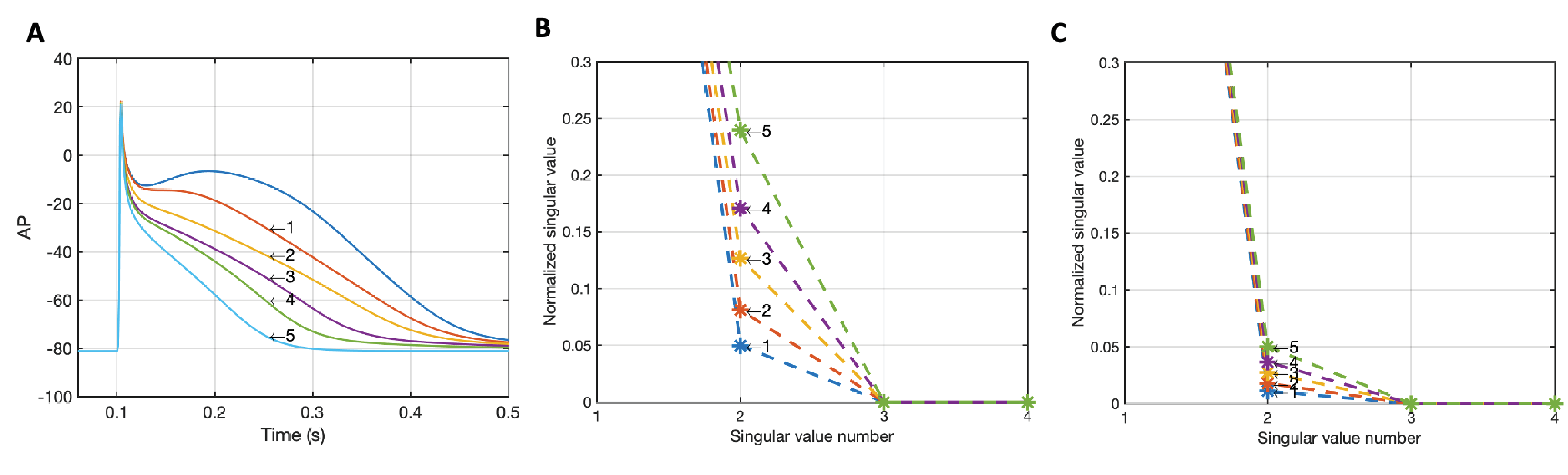}
    \caption{Effect of varying signal morphologies at cell level: A) AP morphologies (activation signals), B) resulting normalized singular values with the same number of cells in group 1 and group 2 C) resulting normalized singular values with different number of cells in group 1 and group 2 .}
    \label{sing_simu_1}
\end{figure*}

   The effect could be calculated in closed form, but is easier appreciated from a simulation. Referring to Fig.\ \ref{sing_simu_1} A), two signal morphologies for the action potential are used: the unmarked blue one, and one of the numbered signals. The signals are derived from the Courtemanche model for a human atrial myocyte, and distinct morphologies are obtained by modifying the calcium current's parameters. A collection of cells are simulated with equal $a_c$, random $\tau_c$, and a specified fraction assigned to each of the two morphologies. In both cases, signals are scaled to have equal ${l_2}$ norm.
   The resulting singular values are shown in Panel B and C (where $\sigma_1$ is normalized to 1, and we zoomed in on the range between 0 and 0.4) . In Fig.\ \ref{sing_simu_1} B), the fraction of cells in either group is equal, while in Fig.\ \ref{sing_simu_1} C), the fraction of cells in either group has a 1:80 ratio.  In Panel B, it is seen that the second singular value increases if the second signal is more different.  In Panel C, it is seen that the differences are more subtle if there is a significant imbalance in the number of cells between the two groups. 

   This result extends to more than two different AP morphologies, but although the number of terms in (\ref{B2}) increases, the columns $\{\bs_i\}$ tend to be parallel and at some point the singular values will not increase by much.

\subsubsection{Electrogram with single AP morphology, flat wavefront} \label{sec:flat}
   Let us now consider the electrode signals. For a single AP morphology $s(t)$, we obtain from (\ref{egm})
   \begin{equation}\label{dm_omega}
   \Tilde{d}_m(\omega) = 
   \sum_{c=1}^{N_c} a_{m,c} \, e^{-j\omega\tau_c}\, \Tilde{s}(\omega) \,.     
   \end{equation}

    Due to the summation over the cells in \ref{dm_omega}, taking the absolute value $|\Tilde{d}_m(\omega)|$ will not have the desired effect of removing the phase factors $e^{-j\omega\tau_c}$. As a consequence, the resulting matrix $\bB$ constructed from $\Tilde{d}_m(\omega)$ will generally have full rank.

    Let us make the simplifying assumption that the atrial wave travels in a single flat wavefront (i.e., a plane wave), and that the coefficients $a_{m,c}$ are spatially invariant except perhaps for an electrode gain, as in (\ref{egm}).  In that case, the phase factors average to
\begin{equation}  \label{flatapprox}
   \sum_{c=1}^{N_c} a_{m,c} \, e^{-j\omega\tau_c}
   \quad=:\quad
   a_m e^{-j\omega \tau_m}
\end{equation}
   where $\tau_m$ is the delay for the cell under electrode $m$. Under this condition, we can again write 
\[
    \bB = \ba \bs^T
\]
   and only one singular value will be nonzero. A proof of this claim is in the Appendix; the proof shows that it does not matter if the electrodes are in a grid or more randomly placed.
   The condition of a single flat wavefront describes the situation of a heart in sinus rhythm (SR), with the activating source sufficiently far away from the electrode.

\subsubsection{Electrogram with single AP morphology, curved or multiple wavefronts}
\begin{figure*}
    \centering
    \includegraphics[width=0.9\textwidth]{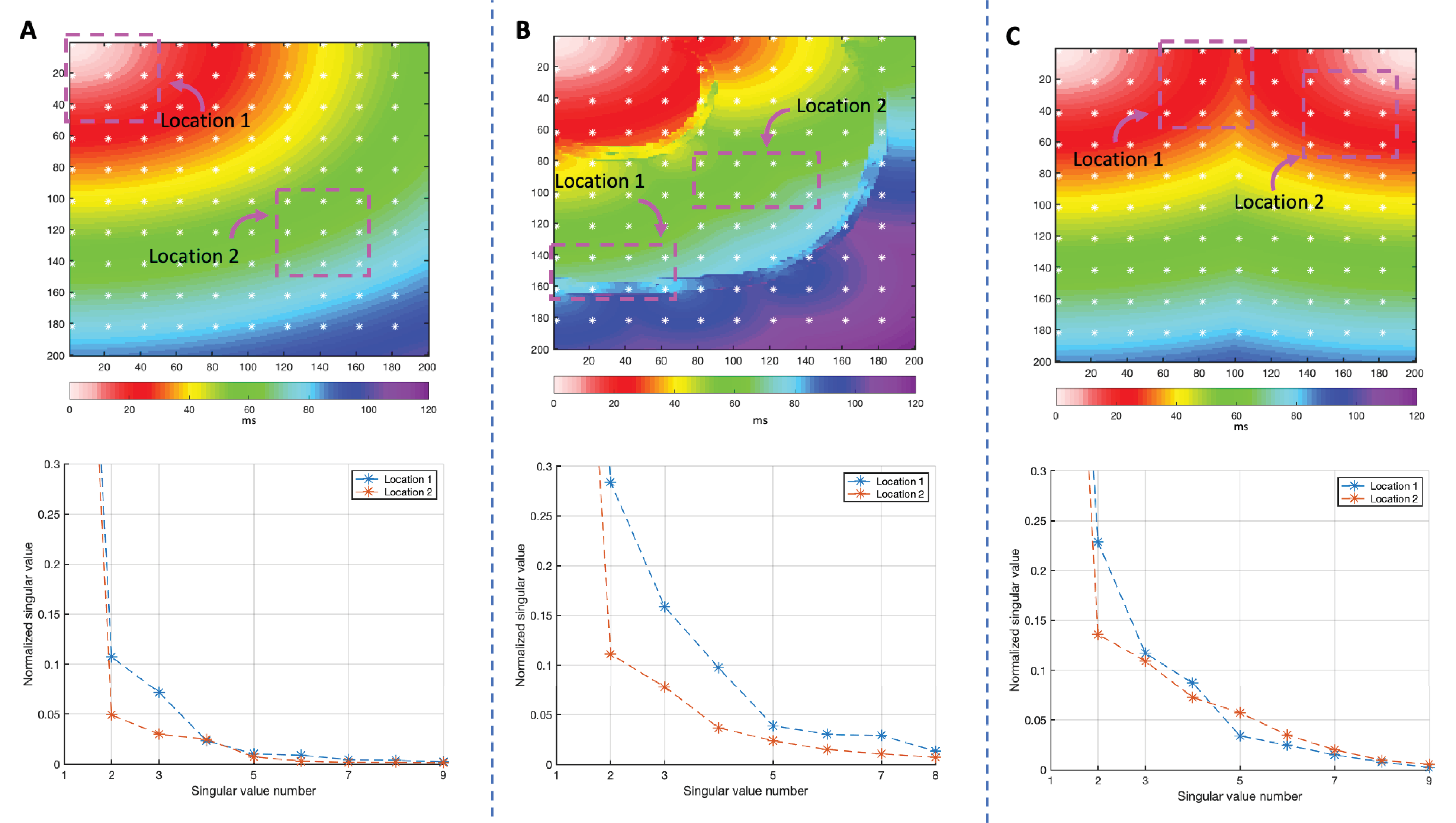}
    \caption{Effect of wavefront curvature: A) single wavefront, B) tissue with conduction blocks C)
    multiple wavefronts with different AP morphologies. \\ Top row: activation map, bottom row: resulting normalized singular values.}
    \label{LAT_sing_sim}
\end{figure*}
   If the wavefront under an electrode is sufficiently curved, then (\ref{flatapprox}) does not hold. As a consequence, more singular values will be nonzero. It is hard to analyze this more quantitatively, but if the delay differences are small, then this effect is not expected to be very strong. 
   A curved wavefront occurs if the activating source is close to the electrode, or in case of a nearby focal activation.  The effect is shown in Fig.\ \ref{LAT_sing_sim} A), where we compare two regions (i.e., $\bB$-matrices) each with 9 electrodes. The tissue is activated in the top-left corner. For Location 1, where the activation wavefront is strongly curved, the second singular value is higher than for Location 2, where the activation wavefront is almost flat. It is also seen that more than two singular values are raised.

   A stronger effect is expected in case an electrode sees multiple wavefronts. This will significantly destroy the symmetry which was a condition to arrive at a rank-1 model. This case relates to the occurrence of fractionation or double potentials, and therefore is associated with AF. Fig.\ \ref{LAT_sing_sim} B) shows the effect. The electrodes in Location 2 see a single wavefront, while the electrodes in Location 1 are above a block and see two wavefronts with clearly different LATs.  In the latter case, the singular values are substantially raised.
  
\subsubsection{Electrogram with multiple AP morphologies}
   If we have cells with two types of morphologies, then (as before) the rank of $\bB$ is increased. If the wavefront is still flat, the presence of two cell types will result in rank 2, but for larger numbers the rank will probably be harder to judge.  If the wavefronts are curved or we have multiple wavefronts, then the number of nonzero singular value will increase as well.

   The effect is shown in Fig.\ \ref{LAT_sing_sim} C), where two signal morphologies are used. Cells activated from the top-left corner use one type of morphology, and cells activated from the top-right corner use a second type. Location 2 has a curved wavefront but only a single signal morphology, which is similar to Fig.\ \ref{LAT_sing_sim} A). Location 1 is a region where the two wavefronts collide and two morphologies are present; this further raises the second singular value.
   
In summary, for the matrix $\bB$ derived from the EGM, we expect a low rank (only one large singular value) in case there is only one AP morphology and the wavefront is flat, a raised second singular value in case there are multiple AP morphologies and/or the wavefront is curved, and a strongly raised second singular value if some electrodes see multiple wavefronts with clearly different LATs. This suggests that the ratio of the second singular value to the first one ($\sigma_2/\sigma_1$) could be a useful feature for detecting and classifying AF. The advantage of this feature would further be that it is directly derived from the data, without relying on the prior estimation of LATs.

The examples further showed that the maximal $\sigma_2/\sigma_1$-ratio that we can expect is about 0.25.

\subsection[Mathematical expression suitable for PDF string]{Definition of a \texorpdfstring{$\sigma_2$ map}}

\begin{figure*}
    \centering
    \includegraphics[width=0.95\textwidth]{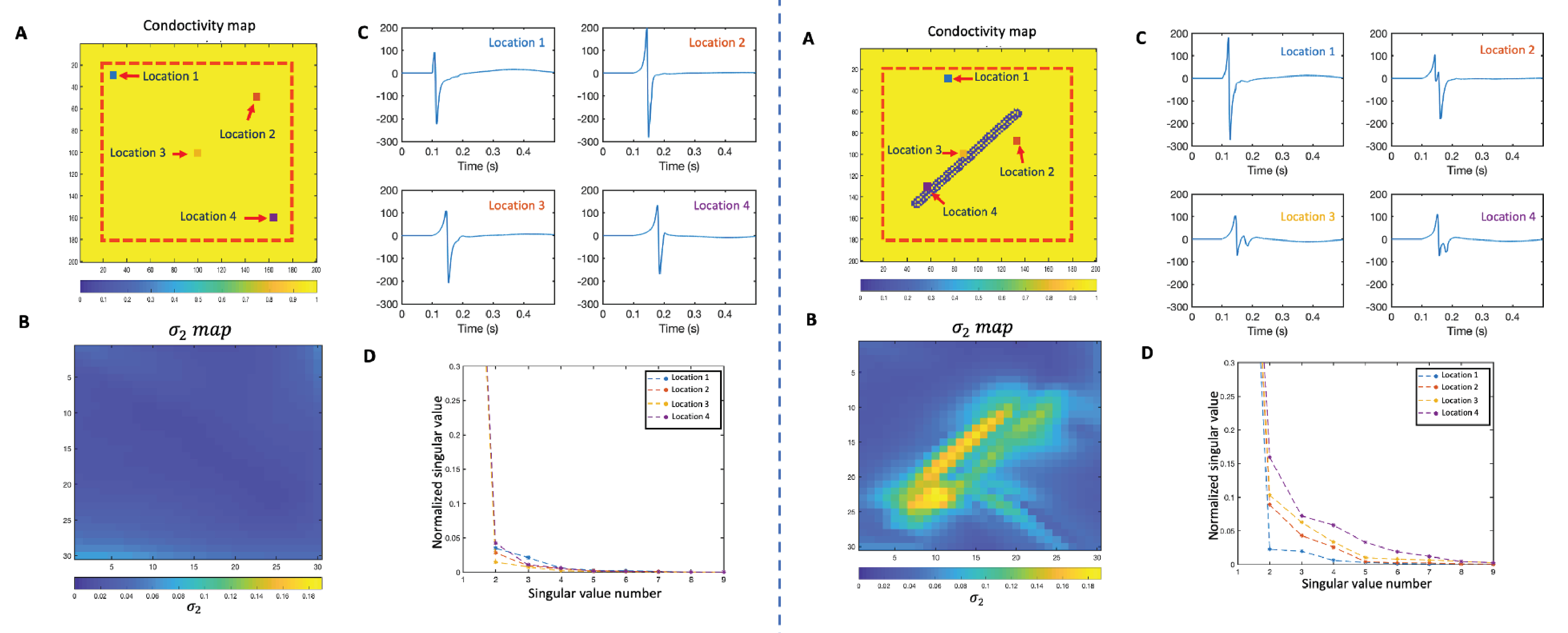}
   \caption{$\sigma_2$ map analysis. Left: homogeneous tissue, right: a tissue with a diagonal line of block. \\
   A) Conductivity map, B) $\sigma_2$ map, C) the electrograms at four electrodes, D) normalized singular values for $3\times 3$ subsets of electrodes, centered at these locations.}
    \label{S2_map}
\end{figure*}

In the presentation of the method, we defined a data matrix $\bB$ obtained from the electrode array. In principle, the matrix could encompass the entire array. The examples presented in Fig.\ \ref{LAT_sing_sim} used subsets of 8 or 9 electrodes, which showed that the normalized singular values vary depending on the location. In locations with a curved wavefront or with multiple wavefronts, the normalized $\sigma_2$ is higher than in locations with a flat wavefront and a single AP morphology. If we collect all electrodes in the data matrix, then the location data is averaged, and the differences between the singular values become smaller and harder to detect.

Therefore, to analyze EGM array data with electrodes arranged in a rectangular grid, we propose to use a ``$\sigma_2$ map'', a location-dependent map. We use $3\times 3$ subsets of the electrode array, and for each subset construct the $\bB$ matrix and compute the normalized $\sigma_2$ value. This gives one pixel in a $\sigma_2$ map, located at the center of the subset. The subset is shifted to cover the entire rectangular array, resulting in the $\sigma_2$ map. 

As an example, Fig.\ \ref{S2_map} shows a simulation where we compare a $\sigma_2$ map for a tissue with a homogeneous conductivity (left part) to a tissue with a conduction block (right part). A rectangular electrode array with $32\times32$ electrodes is placed within the area denoted by the red dashed line. The tissue is activated from the top left and a single AP morphology is used. For the homogeneous tissue, all $3\times 3$ subsets result in normalized $\sigma_2$ values of less than 0.05. For the tissue with a block, the normalized $\sigma_2$ is larger than 0.05 around the block, and is easily recognized in the map. The corresponding time domain signals (e.g., at location 3 and location 4) show double potentials, while in the homogeneous area (e.g., at location 1) we see a single potential. Thus, the $\sigma_2$ map can rapidly point out the inhomogeneities and blocks in the tissue. The advantage of this method is that a LAT estimation and analysis is not required for detecting the blocks.
\subsection{Data}
To evaluate the method's performance and reliability in Sec.\ \ref{sec3}, we use simulated and clinical data. The clinical data is part of the Halt $\&$ Reverse study, approved by the medical ethical committee (MEC $2014-393$), Erasmus University Medical Center, Rotterdam, the Netherlands. 

\subsubsection{Simulation data generation}
\begin{figure}
   \centering
  \includegraphics[width=0.3\textwidth]{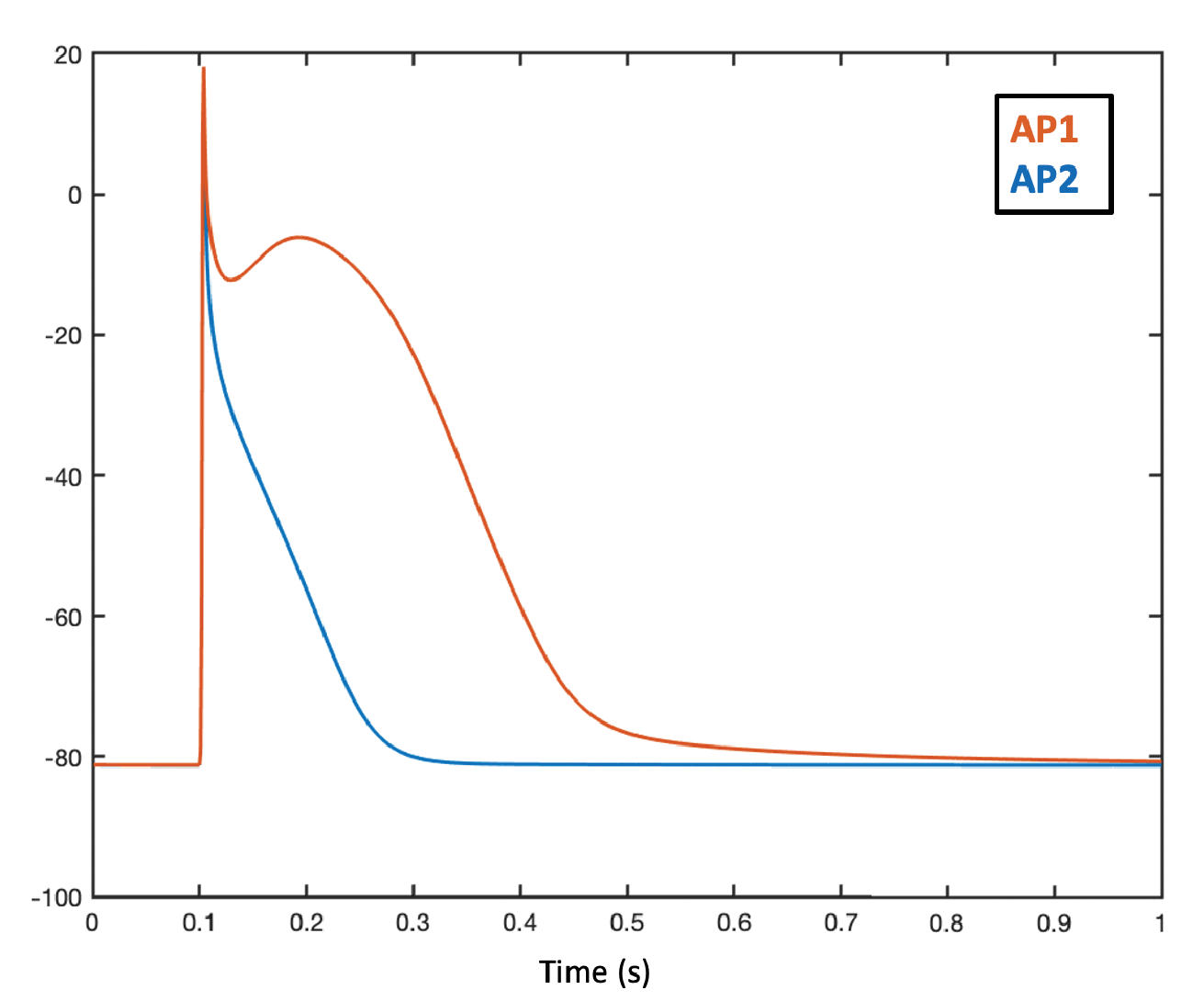}
 \caption{Two simulated morphologies for the action potentials.}
\label{APs}
\end{figure}
We simulated a 2D tissue of $200\times200$ cells. The distance between the cells was $0.1$ mm. We considered two conductivity maps. In the first simulated tissue, we have taken into account a homogeneous tissue where the conductivity is constant throughout the tissue. For the second simulated tissue, two different conductivities have been used: specific cells have a constant conductivity of $c_1=1$, whereas the others have a constant conductivity of $c_2=0.01$.  

The AP signals of the cells are generated at a resolution of 1 kHz, using the Courtemanche model as implemented in \cite{jordi,code}. 
We generated APs with two different morphologies, which are shown in Fig.\ \ref{APs} as AP1 and AP2.
For visualization purposes, activation maps are generated by detecting the activation time as the instant when a cell crosses a threshold of $-40$ mV during the depolarization phase of the AP.

To activate the tissue, two wavefront directions have been used. The first wavefront originates from the top-left corner, and the second is from the top-right corner. The electrogram signals were observed by a rectangular electrode array of $10\times10$ electrodes with inter-electrode distance of $2$ mm, at a constant height of $z_0 = 1$ mm from the tissue. 
To increase the resolution in the  $\sigma_2$ map analysis, we increased the number of electrodes in the electrode array to $32\times 32$ electrodes with an inter-electrode distance of $0.5$ mm.

\subsubsection{EGM data collection}
High-resolution epicardial unipolar EGM data was collected at the Erasmus Medical Center (EMC) during open-heart surgery on patients without a history of AF, as described in more detail in \cite{yaksh2015novel}. Fig.\ \ref{block} A) shows the standardized 9 recording locations; at each location, a recording consists of 5s of SR followed by 10s of induced AF.  The electrograms were recorded using a rectangular electrode array with $8\times 24$ electrodes where the inter-electrode distance was 2 mm and the electrode diameter was 0.45 mm. The signals were amplified, filtered to a frequency range between 0.5 and 400 Hz, sampled at a rate of 1 kHz with resolution of 16 bits, and stored. One lead was used to record the ECG.

During data analysis, we filtered the signals using a Butterworth band-pass filter in the frequency range between 0.33 Hz and 30 Hz \cite{moghaddasi2022classification,chang2010arrhythmia,moghaddasi2021tensor}. The Pan-Tompkins R-peak detection method \cite{pan1985real} was used on the ECG lead to segment the atrial activity of each EGM. To select the atrial activity, we used a fixed window with a length of 260 ms and select the interval between 320 ms and 60 ms before the R-peak \cite{moghaddasi2022novel}. This resulted in $N=130$ frequency-domain samples per beat.

For the EGM data analysis, we included five patients without a history of AF. We pre-screened the available heartbeats using these exclusion criteria: 1) electrically silent heartbeats; 2) heartbeats where the fibrillatory waves are absent in the fixed window. As a result, the number of patients per location varies between 3 and 5, where between 2 and 23 heartbeats are included per patient. In total, 189 and 395 heartbeats are used for SR and AF episodes, respectively. More details about the number of heartbeats per location are reported in Table \ref{tbl1}.

\begin{table}
\centering
\caption{\label{tbl1} Number of selected heartbeats per location}
\begin{tabular}{c c c} 
 \hline
 Mapping location & $\#$ SR heartbeats & $\#$ AF heartbeats \\ [0.5ex] 
 \hline
 BB0 & 19 & 28  \\ 
 LA1 & 22 & 51  \\
 LA2 & 17 & 38  \\
 PVL1 & 20 & 19 \\
 PVR1 & 24 & 35  \\
RA1 & 23 & 41  \\ 
 RA2 & 22 & 60  \\
 RA3 & 22 & 58  \\
 RA4 & 20 & 65 \\
 \hline
\end{tabular}
\end{table}

\subsubsection{Body Surface Potential data collection}\label{bsp}
\begin{figure}
    \centering
    \includegraphics[width=0.4\textwidth]{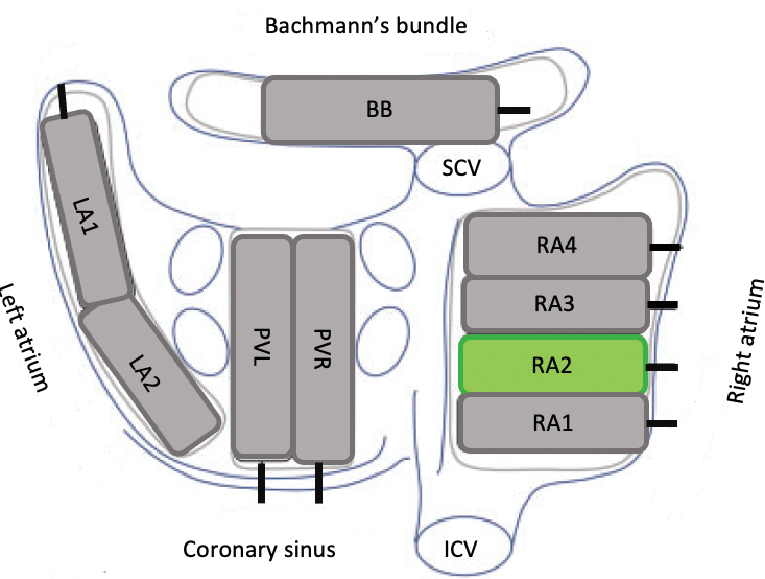}
    \includegraphics[width=0.4\textwidth]{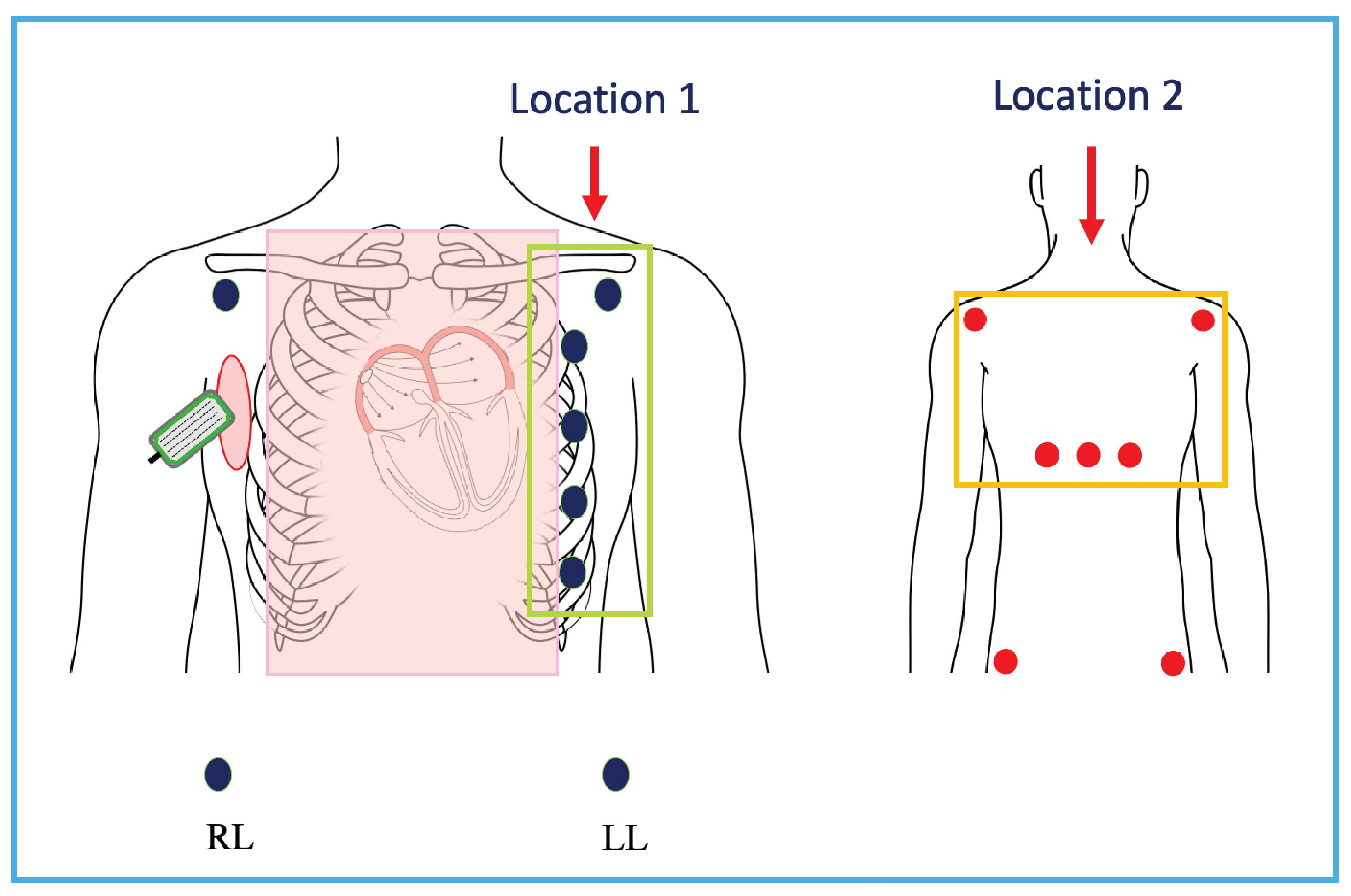}
    \caption{(Top) EGM measurement locations from the posterior view of the atria. (Bottom) Setup to simultaneously measure the EGM at location (RA1, RA2 or RA3) and body surface potentials with 15 electrodes, during minimally invasive surgery. Location 1 and 2 are designated subsets of 5 electrodes on the front and back of the patient, respectively.\\We recorded data from all 9 standardized locations (shown on top) for the EGM data collection, and for the combined EGM and BSP measurements, we recorded at RA1-RA3.\\
    Right atrium (RA), left atrium (LA), right pulmonary vein (PVR), left pulmonary
vein (PVL), Bachmann’s bundle (BB), superior caval vein (SCV), inferior caval vein (ICV).}
    \label{block}
\end{figure}
To be able to simultaneously measure high-resolution epicardial EGMs and multi-lead body surface potentials (BSPs), we designed a novel sub-vest to record the BSPs during minimally invasive surgery. 
We placed the 15 electrodes of the vest at the locations indicated by the circles in Fig. \ref{block}, where black circles denote the electrodes on the front and red circles denote the electrodes on the back of the patient. This design was motivated as follows. First, the area highlighted in the faded color, called the sterile field, is inaccessible during minimally invasive surgery. Second, the atrial activity is a focus of this investigation. Since the atrial activity is generated during the depolarization phase, we covered an area that captures this, i.e.\ optimized for a heart axis between -30 and +90 degrees. Additionally, to capture the atrial activity from the back of the patient, we positioned three electrodes close to the atrium on the back. For practical reasons, we had to limit the total number of leads. We used 15 prewired disposable electrodes from Nissha Medical Technologies (NMT), with the code CLARAVUE 4009839C, which are attached to the patient before starting the surgery. A 6-7 cm incision called an auxiliary port is made in the third or fourth intercostal space to perform the surgery. After positioning the electrode array to the right atrium at three locations marked RA1, RA2, and RA3 in Fig. \ref{block}, high-resolution electrograms and multi-lead BSPs were simultaneously measured.

We acquired measurements from 1 male patient without a history of AF. The patient underwent mitral valve prolapse (MVP) surgery. The left ventricular ejection fraction (LVEF) was normal, and the body mass index (BMI) was 26.5.  We recorded for 30s during SR and 30s during an induced AF episode. A similar filtering and segmentation approach has been implemented for the BSP measurements.

\section{Results}      \label{sec3}
From Sec.\ \ref{sec2}, the hypothesis is that the normalized $\sigma_2$ value is a useful indicator to detect and classify cases of AF. This is tested on both simulated and two types of clinical data. We used the entire data matrix to demonstrate the general characteristics of AF by the normalized singular values. While using submatrices to locate the electropathological areas in the tissue allowed us to learn more about the spatial distribution of the AF substrate in the tissue. 

\subsection{Simulation results}

\begin{figure}
    \centering
    \includegraphics[width=0.45\textwidth]{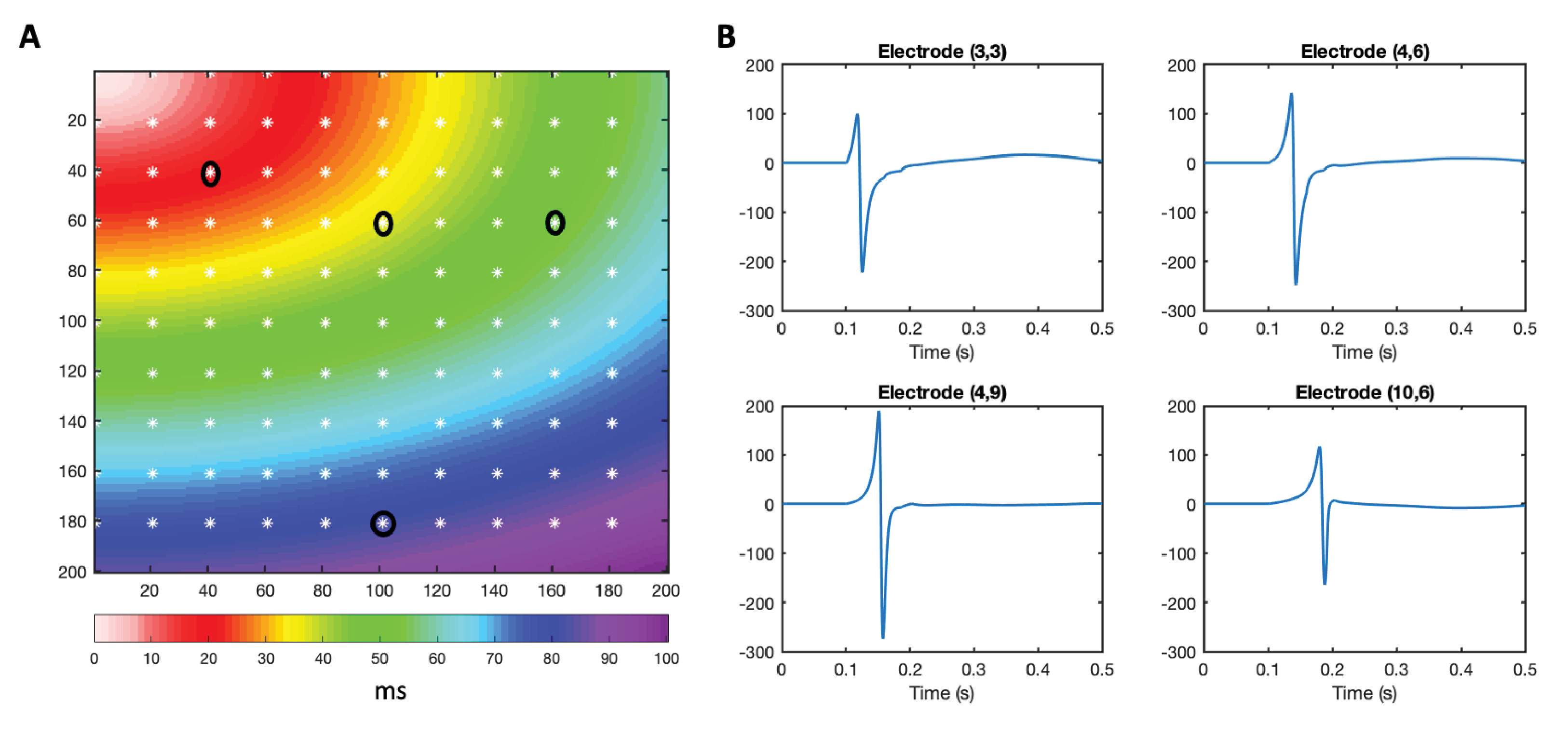}
    \caption{One wavefront, one AP morphology: activation map, and example electrograms.}
    \label{Actmap_1s}

    \centering
    \includegraphics[width=0.45\textwidth]{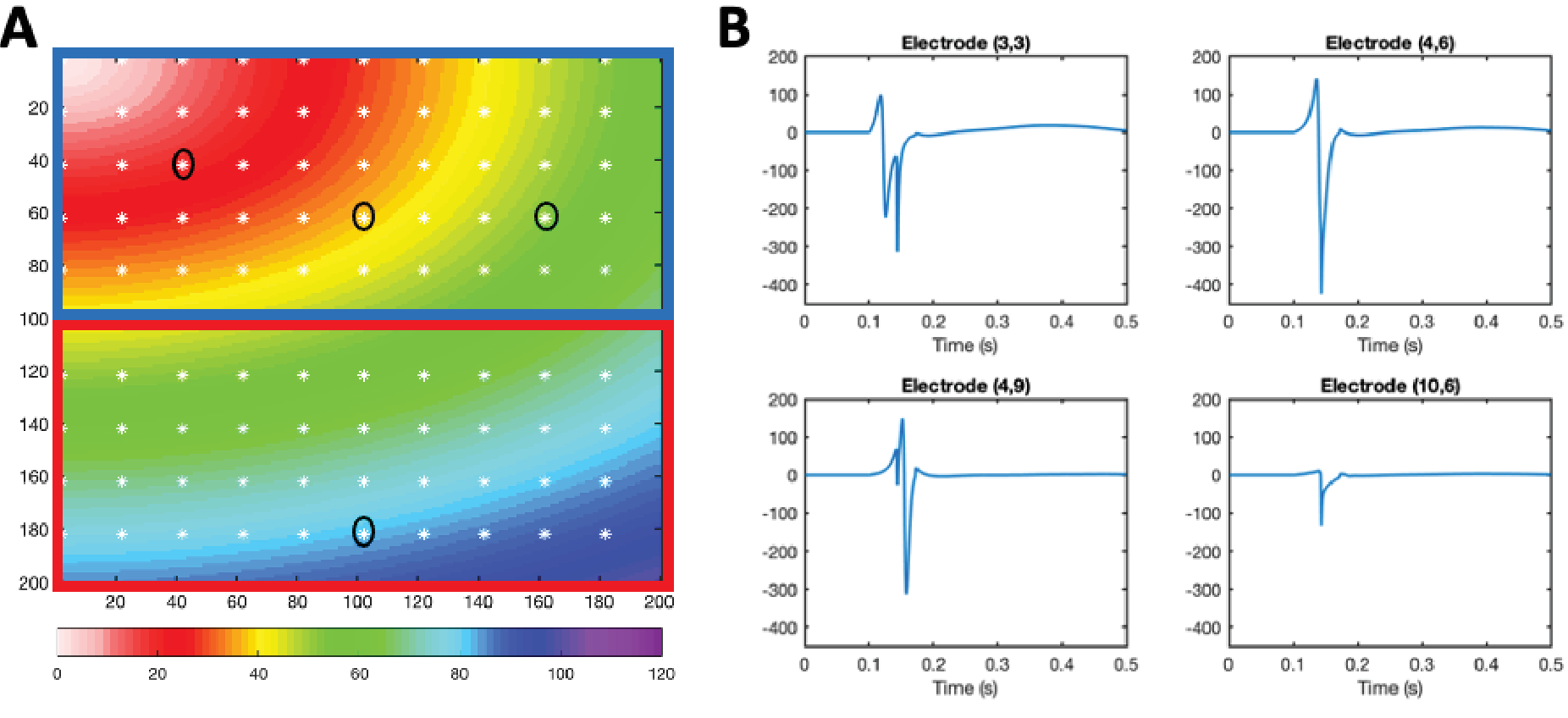}
    \caption{One wavefront with two conduction areas with different AP morphologies: activation map and electrograms.}
    \label{Actmap_1s2ap}
    
    \centering
    \includegraphics[width=0.45\textwidth]{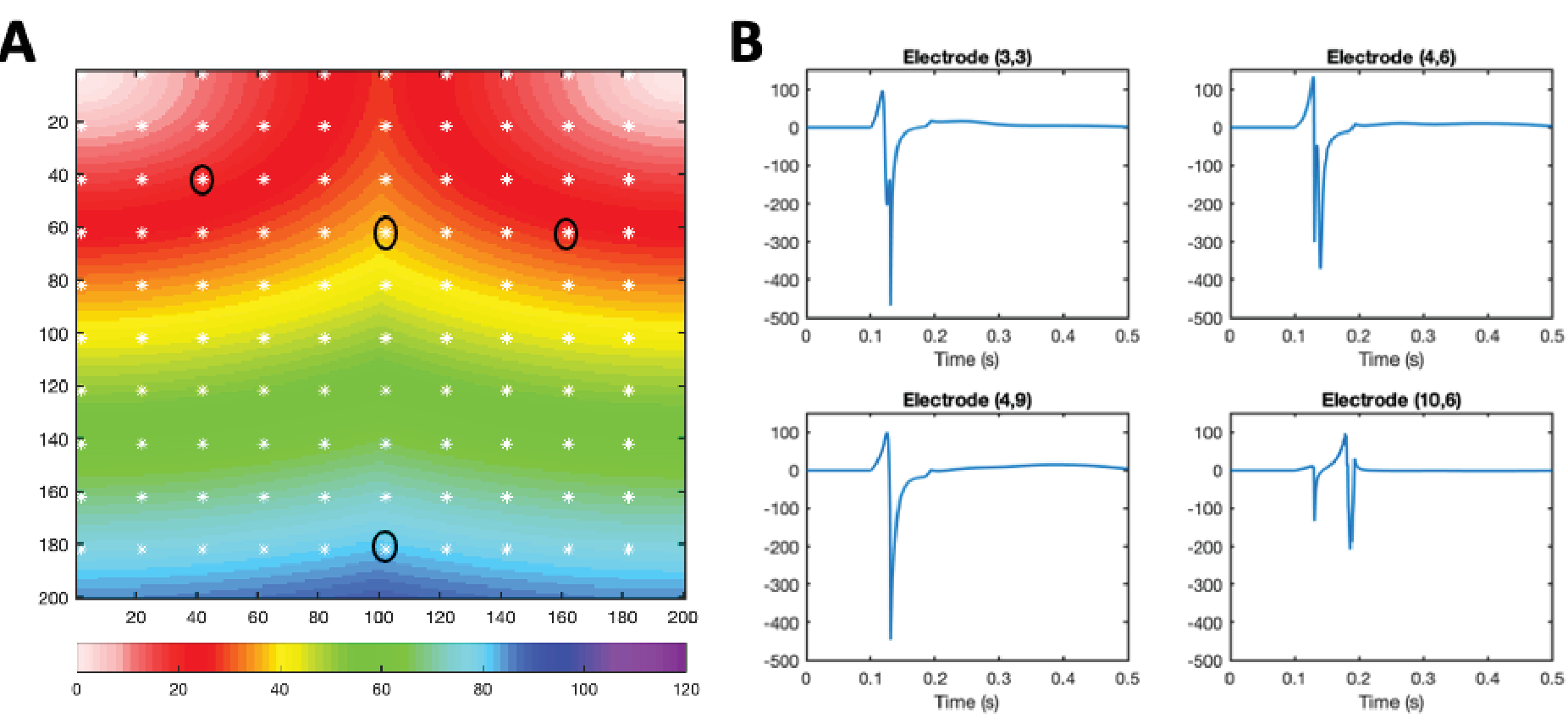}
    \caption{Two wavefronts with two different AP morphologies: activation map and electrograms.} 
    \label{Actmap_2s}

    \centering
    \includegraphics[width=0.45\textwidth]{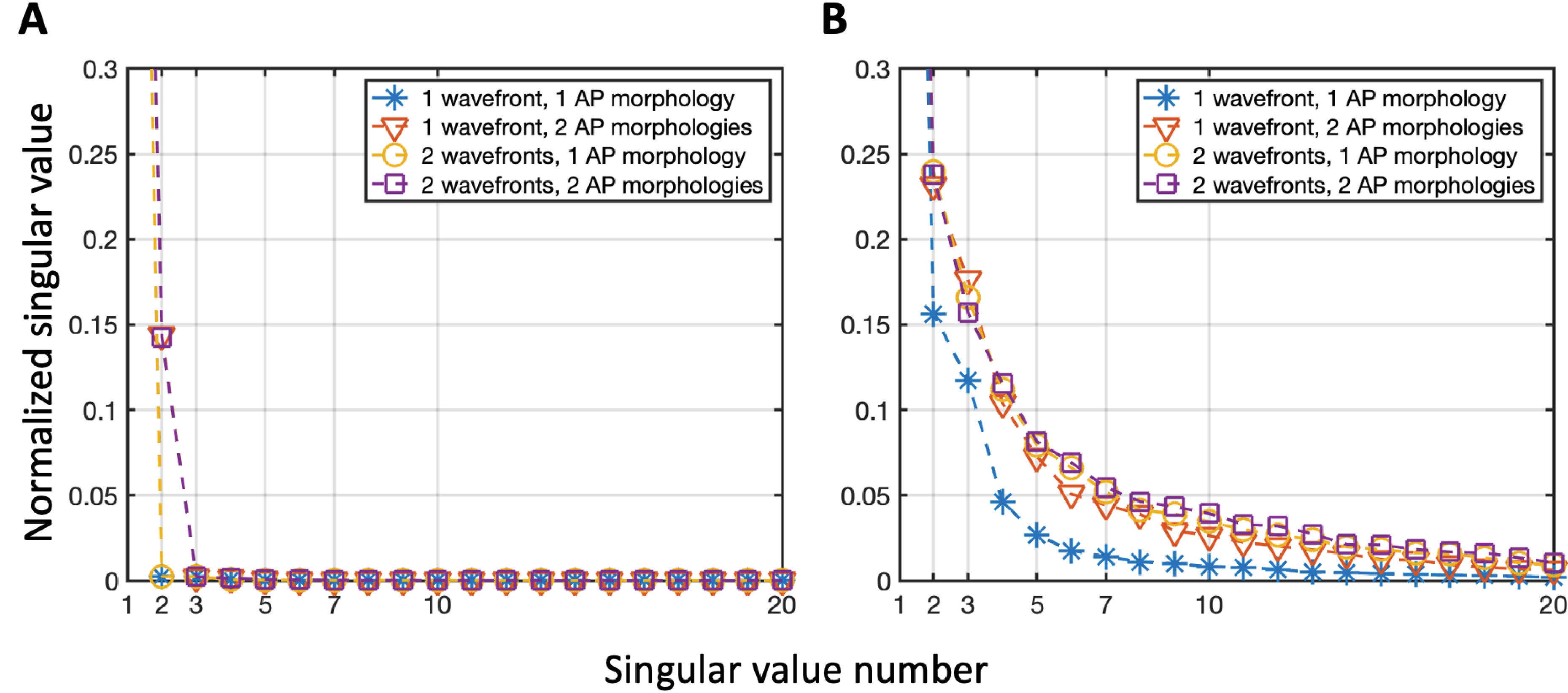}
    \caption{Simulated data: normalized singular values using the entire data matrix. A) cell level, B) epicardial level. }
    \label{sing_simu}
\end{figure}

\subsubsection*{Setup}
   We compare three cases. In the first case, we generated a homogeneously conducting 2D tissue in which the AP is initiated from the top-left corner and propagates throughout the tissue.   The activation map and electrograms at a few selected electrodes are shown in Fig.\ \ref{Actmap_1s}. As can be seen, the electrograms have similar morphologies, with a single deflection (a so-called  single potential). 

   For the second case, we generated one wavefront with two different AP morphologies. The activation map and selected electrograms are shown in Fig.\ \ref{Actmap_1s2ap}; the areas with two different morphologies are indicated by rectangles. Fractionated potentials can be seen in the electrograms.

   For the third case, we generated two wavefronts, activated from the top-left and top-right corners of the simulated tissue. The conductivity was homogeneous, and the two wavefronts collided on the center axis. The cells at the left and the right generated different AP morphologies AP1 and AP2 (Fig.\ \ref{APs}). Fig.\ \ref{Actmap_2s} shows the activation map and example electrograms. Some electrograms are fractionated, and this is more pronounced at the locations of wave collisions.

\subsubsection*{Results}
   The three scenarios are first tested at the cell level. The results are in Fig. \ref{sing_simu} A). The figure shows that the normalized singular values of the $\bB$-matrix are increased only by the variety of AP morphologies, but not by the number of wavefronts. This is because at the cell level, the $\bB$-matrix is not sensitive to activation time.

   However, at the epicardial level (Fig.\ \ref{sing_simu} B), it is seen that the normalized singular values of the $\bB$-matrix are increased in all cases. For the case with 1 wavefront and 1 AP morphology, this is because the wavefront is curved. For the other cases, it is a combination of the curved wavefronts and the variation in the AP morphology. Also with a single AP morphology, the fact that there are 2 colliding wavefronts will increase the singular values. These effects are apparently not additive: the first scenario gives a normalized $\sigma_2$ value of 0.15, the other scenarios result in 0.24.

\subsection{Clinical results}
\begin{figure*}
    \centering
    \includegraphics[width=0.9\textwidth]{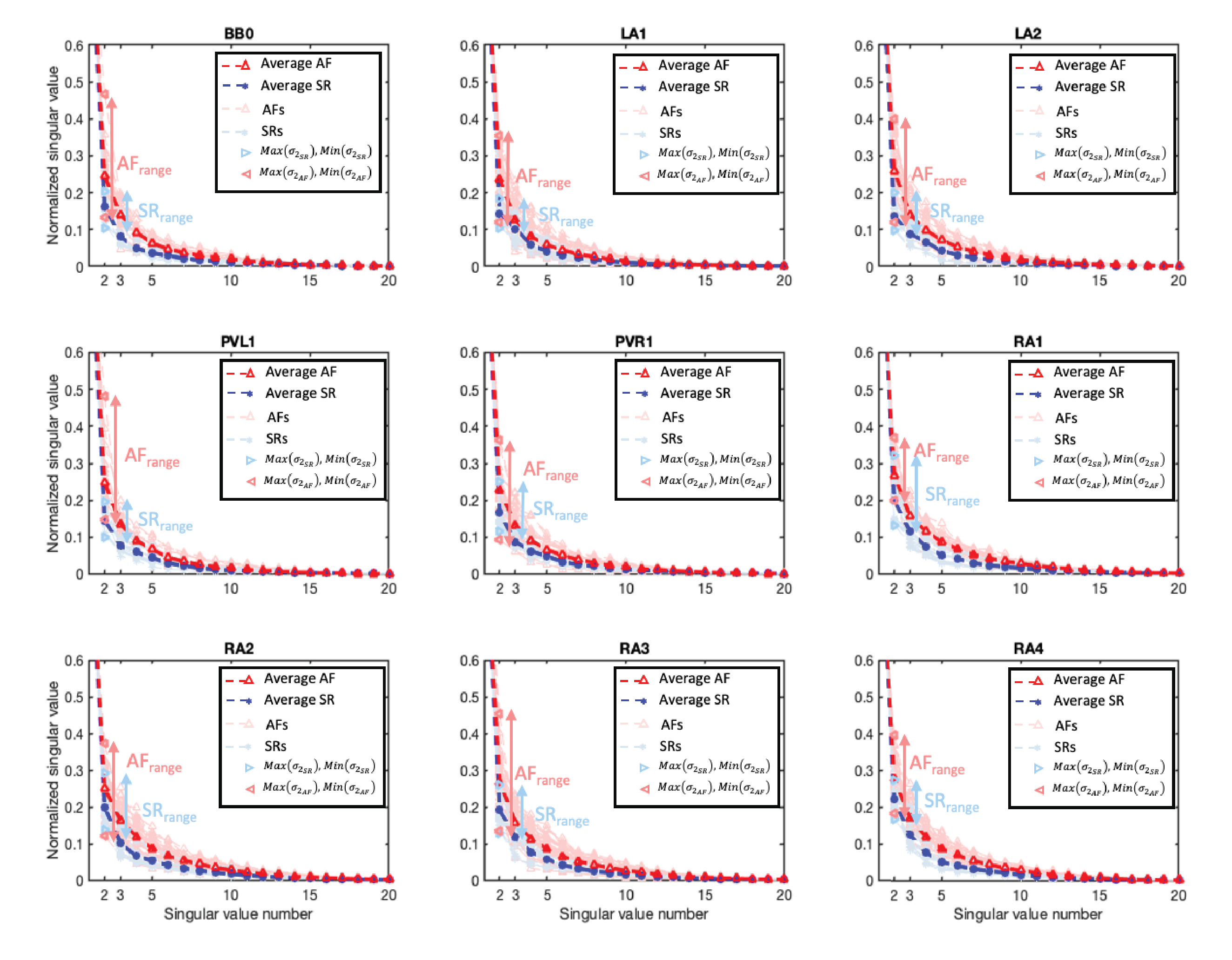}
   \caption{EGM clinical data: Normalized singular value analysis for each heartbeat per location. The bold lines show the average.}
    \label{s_val_9loc}
\end{figure*}
\begin{figure}
    \centering
    \includegraphics[width=0.5\textwidth]{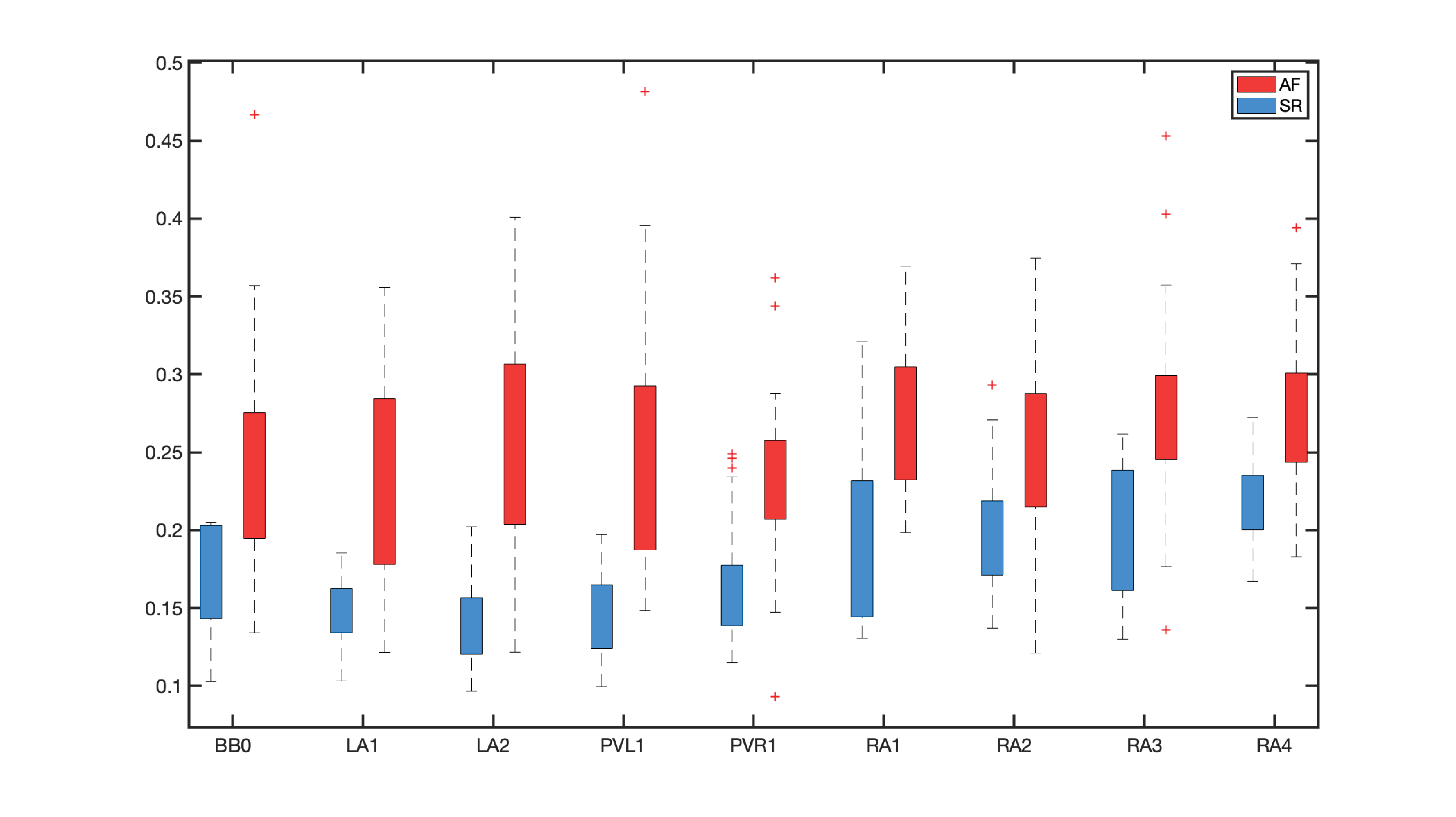}
    \caption{EGM clinical data: Box plot showing the distribution of the normalized $\sigma_2$ over the heartbeats for each of the mapping locations. Within the box, values between the 25th and 75th percentiles are shown.}
    \label{boxplot}
\end{figure}

We applied the proposed method, normalized singular values, to the clinical EGM data as well. Fig. \ref{s_val_9loc} shows in blue and red faded lines the normalized singular values for each heartbeat during the SR and AF episode per location. The averages over the heartbeats are shown in bold blue and red. It is evident that the singular values are higher during AF heartbeats than during SR heartbeats. Arrows show the range of normalized $\sigma_2$ values, which is smaller for SR than for AF. 

The distribution of the normalized $\sigma_2$ over the heartbeats is shown more clearly in Fig.\ \ref{boxplot}. It can be seen that there is a significant difference between the normalized $\sigma_2$ during SR and the normalized $\sigma_2$ during AF (P-value $<$ 0.001 for all mapping locations). These results show that the normalized $\sigma_2$ is a helpful feature to discriminate between SR and AF.  The threshold to separate the distributions appears to be location specific.

\subsection{Combined EGM and Body Surface Potential data results}

\begin{figure}
    \centering
    \includegraphics[width=0.45\textwidth]{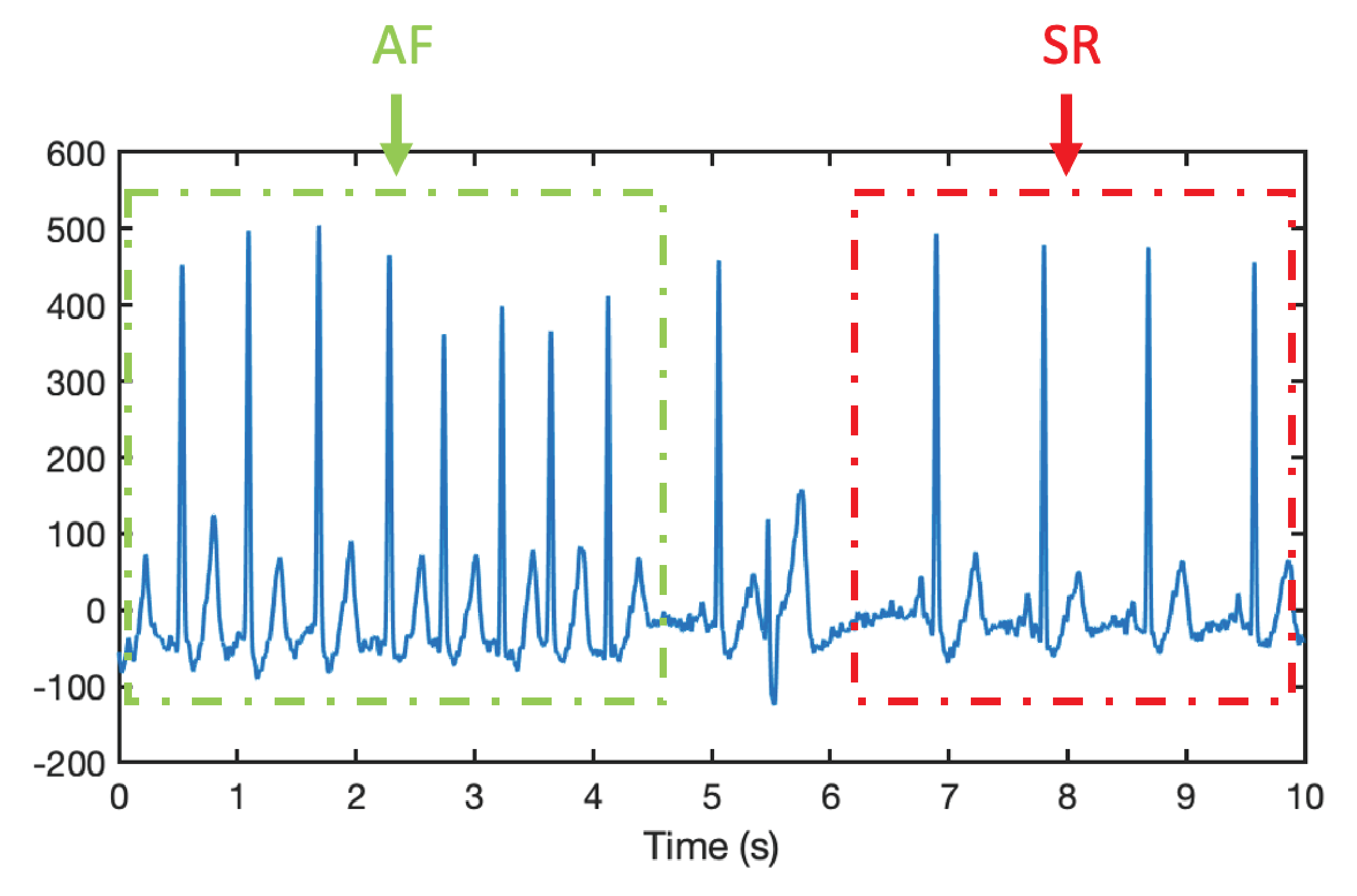}
    \caption{One lead from the BSPs containing AF and SR heartbeats.}
    \label{ECG_AF_SR}

    \centering
    \includegraphics[width=0.5\textwidth]{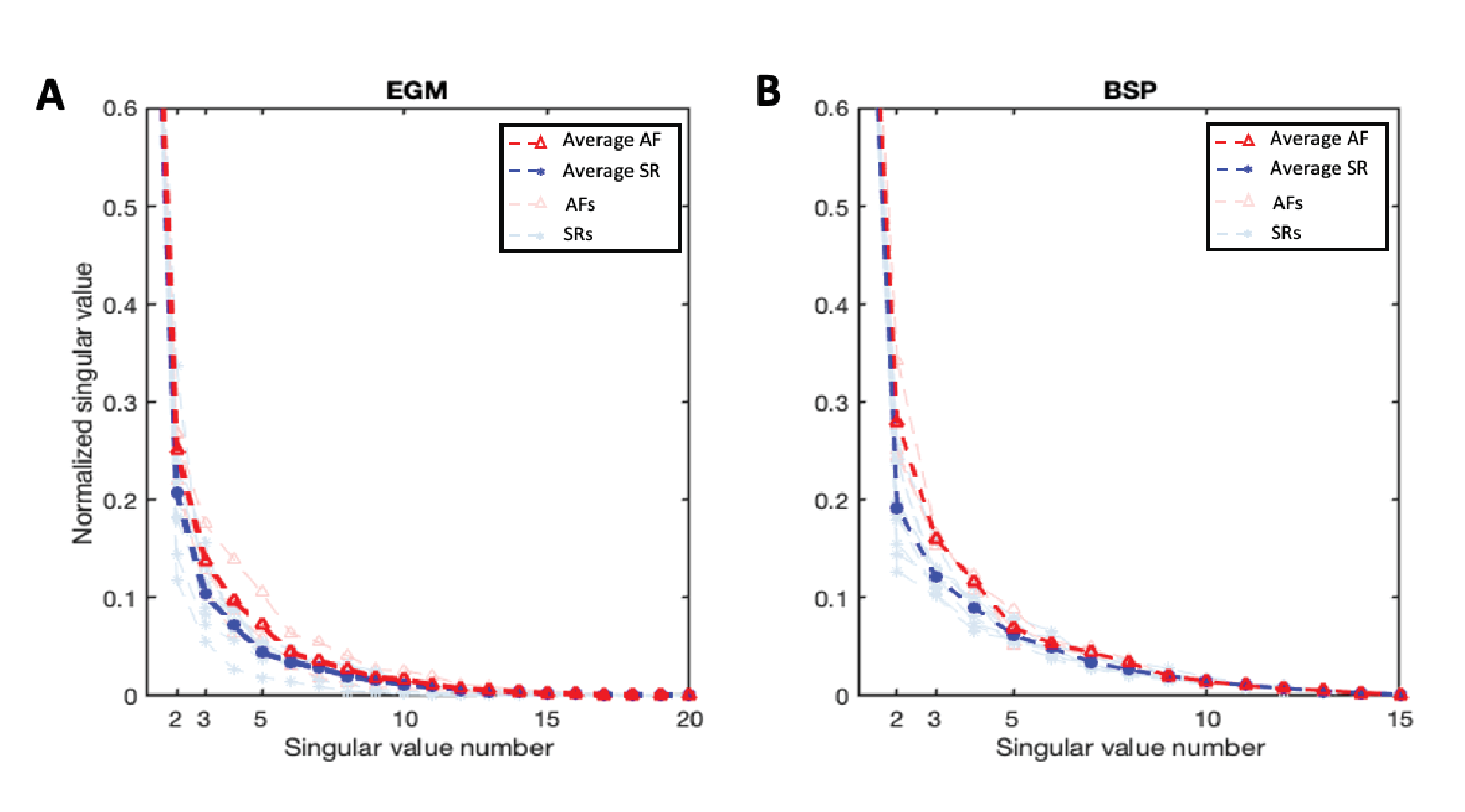}
    \caption{Singular values of each heartbeat during SR and AF episodes using simultaneous EGM and BSP measurements. A) EGM data at the RA1 location, B) BSP data. The bold lines show the average.}
    \label{EGM_BSP}

    \centering
    \includegraphics[width=0.45\textwidth]{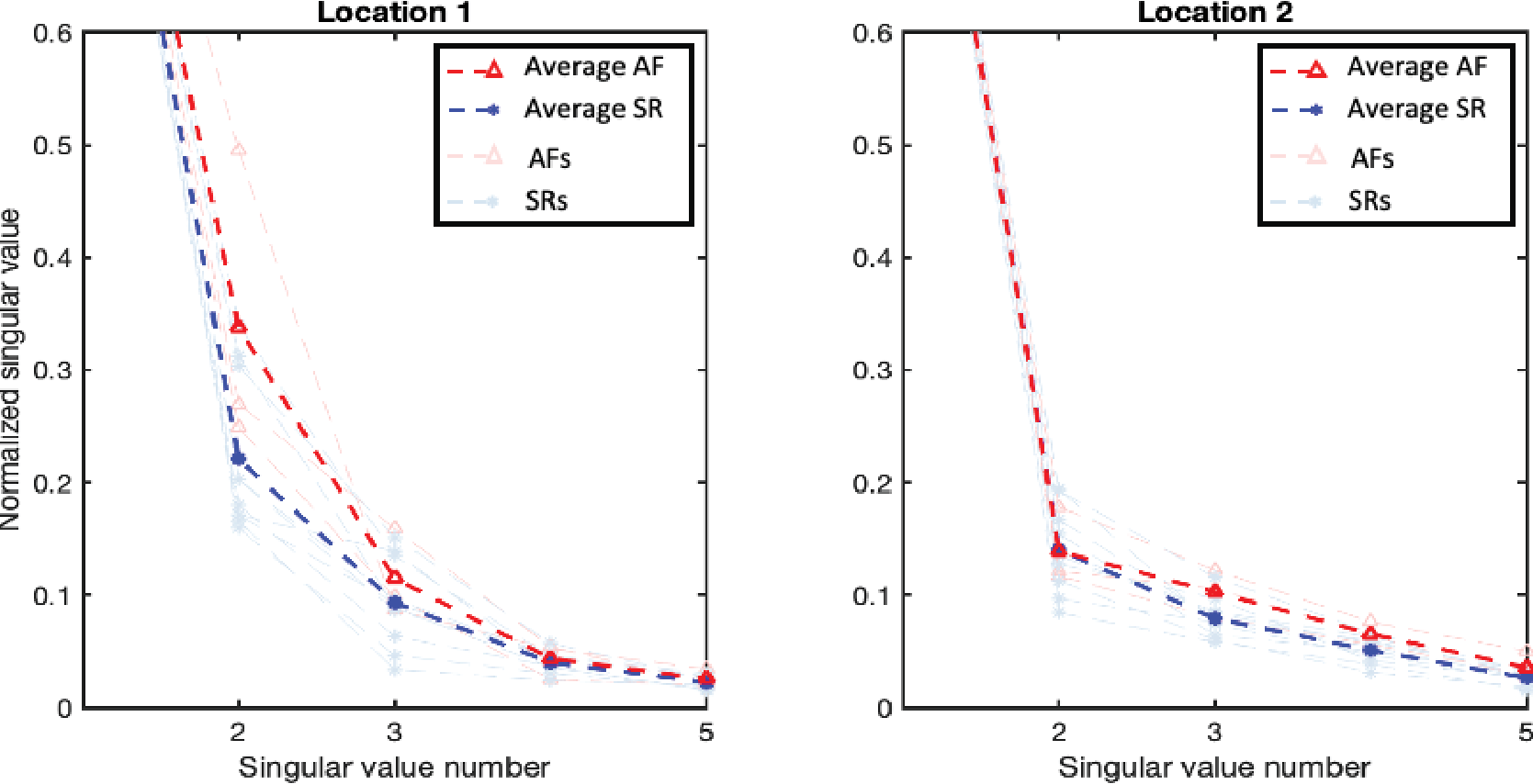}
    \caption{Effect of BSP electrode location on the normalized singular values for two selected subsets of 5 electrodes. }
    \label{BSP_L1_L2}
\end{figure}

For the combined EGM and BSP data, we investigated the singular values of the $\bB$ matrix during SR and induced AF on a single patient without history of AF. Fig.\ \ref{ECG_AF_SR} shows a segment of one lead of the measured data where the patient was in AF for 5s and then the rhythm returned by itself to the normal sinus rhythm. The EGM and BSP data has been measured simultaneously. The normalized singular values for the SR and AF heartbeats for both EGMs and multi-lead BSPs are shown in Fig.\ \ref{EGM_BSP}, using the data at location RA1. The average singular values across the heartbeats are shown in bold blue and red for the SR and AF beats, respectively. It can be seen that the singular values of the heartbeats during AF are higher than during SR. This difference is somewhat more pronounced for the BSP data than for the EGM data: the average normalized $\sigma_2$ is 0.29 during AF and 0.19 during SR.

To study this in more detail, we sub-divided the BSP data to investigate the location dependency of the singular values. Referring to Fig.\ \ref{block} and Fig. \ref{BSP_L1_L2}, we took 5 electrodes in the anterior plane, denoted by the green box ({\em location 1}), and 5 electrodes in the posterior plane, denoted by the orange box ({\em location 2}). At location 1, the average singular values of the AF beats are strongly higher than the average singular values of the SR beats ($\sigma_2$ at 0.35 vs.\ 0.22). In contrast, at location 2, the average singular values of the SR and the AF beats are similar ($\sigma_2$ at 0.15). This demonstrates that the presence of abnormal regions is not visible in all electrodes, and a selected subset might show a stronger response than the full data matrix.

\begin{figure}
    \centering
    \includegraphics[width=0.5\textwidth]{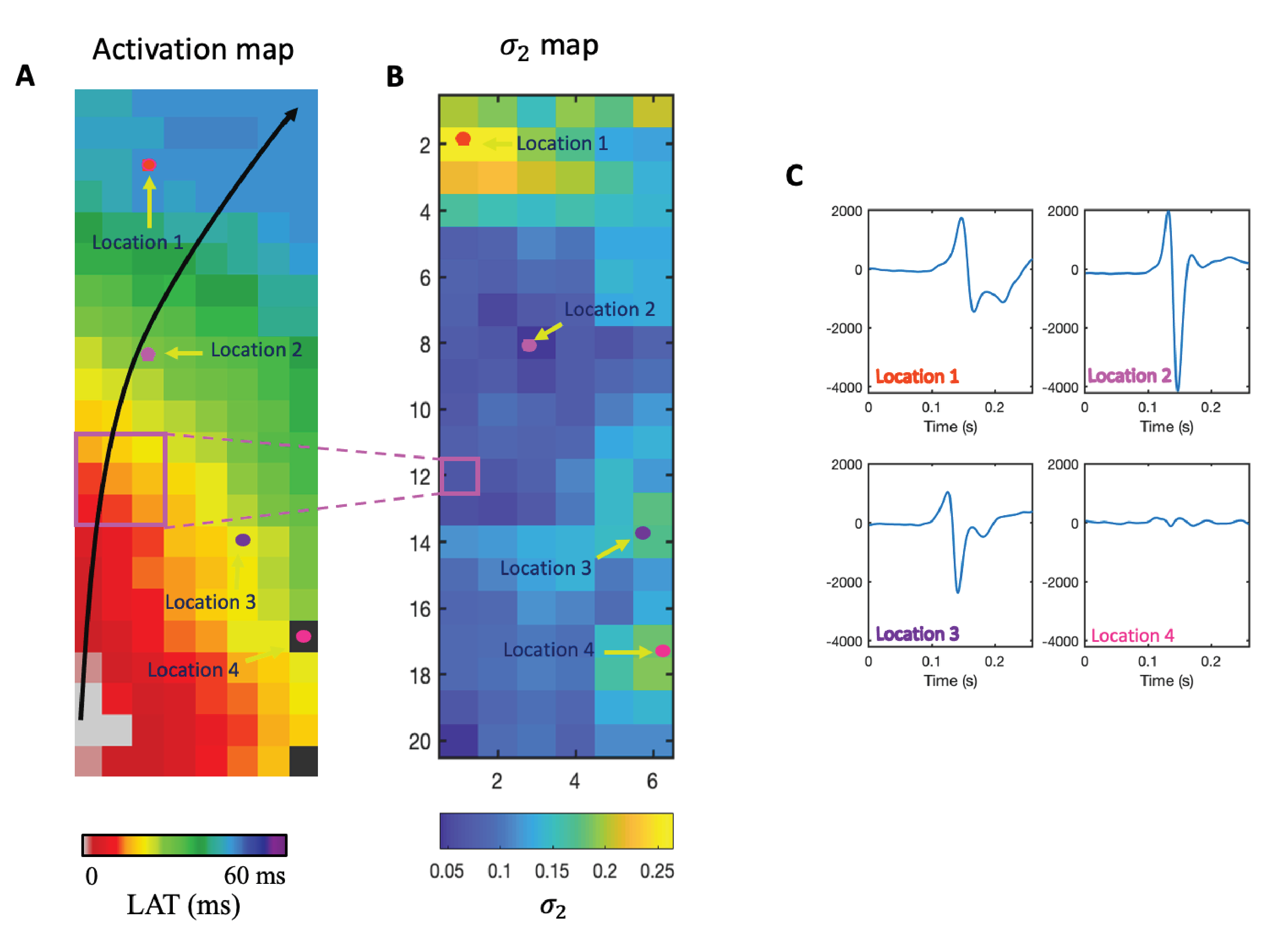}
    \caption{$\sigma_2$ map analysis in SR, A) activation map, B) $\sigma_2$ map, C) EGMs at four electrodes. The black arrow shows the wavefront propagation. }
    \label{s2_sr_clinical}

    \centering
    \includegraphics[width=0.5\textwidth]{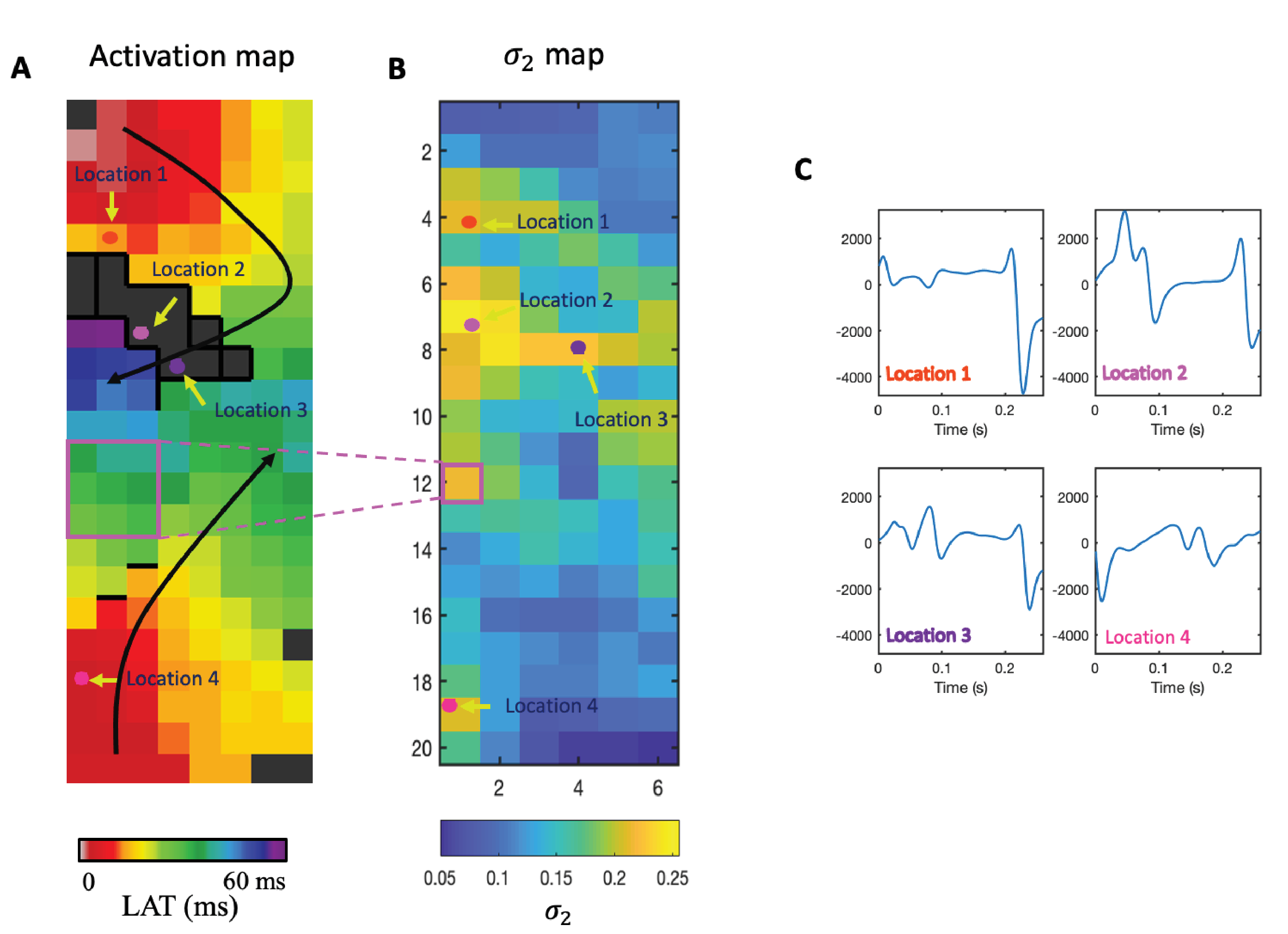}
    \caption{$\sigma_2$ map analysis in AF, A) activation map, B) $\sigma_2$ map, C) EGMs at four electrodes. The black arrows show the wavefront propagation.}
    \label{s2_af_clinical}
\end{figure}

The EGM data at the location RA1 can be studied in more detail using the activation map and the $\sigma_2$ map. A single heartbeat during SR is shown in Fig. \ref {s2_sr_clinical} A. The wave propagation starts from the bottom-left corner, denoted by grey in the activation map. The wavefront propagation is shown by the black arrow, which is analysed by a qualified physician. The wavefronts appear to be flat (linear) in most of the map. Using an overlapping $3\times3$ electrode array (e.g., the area indicated by the pink square), we construct the $\sigma_2$ map shown in Fig. \ref {s2_sr_clinical} B). Generally, the normalized $\sigma_2$ is less than 0.1 (blue color), except for a few regions. It demonstrates that on the areas with a flat wavefront, the rank 1 approximation in Section \ref {sec:flat} holds. However, at location 1, the normalized $\sigma_2$ is about 0.25 (yellow color). The corresponding EGM in Fig.\ \ref {s2_sr_clinical} C) shows that location 1 has a double potential. This abnormality is not seen in the activation map. In other words, while the LATs remain normal, the $\sigma_2$ map visually indicates areas with altered morphology.

The activation map for a single heartbeat during the induced AF episode is shown in Fig.\ \ref {s2_af_clinical} A). It can be seen that the tissue under the electrode array starts to be activated at multiple locations (bottom left and top left). Furthermore, some areas have conduction blocks (CB), as marked by the bold black lines. A CB is declared when the LAT difference between two adjacent electrodes is greater than or equal to 12ms \cite{ye2021signal,lanters2017spatial}. The $\sigma_2$ map for the same heartbeat is shown in Fig.\ \ref {s2_af_clinical} B). At the area with CB around locations 1, 2 and 3, we observe that $\sigma_2 \geq 0.2$ (orange/yellow color), while in the areas where the wave propagates normally, the $\sigma_2$ is less than 0.1 (blue color). Looking at the corresponding EGM examples in Fig.\ \ref {s2_af_clinical} C, at locations 1-3 we observe signals with a double potential followed by a single potential. It demonstrates that the double potential regions can be detected by the $\sigma_2$ map. Further, at location 4, the activation map shows normal wave propagation, while the $\sigma_2$ is greater than 0.2 (orange color). The EGMs at location 4 shows a single potential followed by a double potential. Thus, the $\sigma_2$ map can point at double potentials in some regions while these are not visible in the activation map. Conversely, the activation map shows some CB (black lines) above location 4, while the $\sigma_2$ map gives no trigger at this location.

\section{Discussion}\label{sec4}
Summarizing the results, we have shown that the normalized $\sigma_2$ of the $\bB$-matrix from subsets of electrode data is sensitive to curved wavefronts, conduction blocks, and variations in AP morphology. These changes can be detected in EGMs and also in multi-lead ECGs, if the electrodes are positioned at favorable locations. Further, we have seen that 
the $\sigma_2$ map is a useful tool for detecting such changes, complementary to the use of activation maps. 

An increased normalized $\sigma_2$ is often related to the occurrence of double potentials or simultaneous presence of multiple AP morphologies in the considered subset of electrodes. These are often associated with AF. In the clinical data, we showed that the heartbeats annotated as AF always scored higher than heartbeats in SR.

If we assume that AF initiation and progression can be modeled by a variation in the morphology of APs, then the normalized $\sigma_2$ can be a useful feature to detect such changes. 
The array data can be processed for each individual heartbeat, and by tracking the normalized $\sigma_2$, we can efficiently monitor the evolution of a patient over time.
In preliminary research, we have shown that variations of this feature are able to discriminate between paroxysmal (short-lasting AF) and persistent (long-lasting AF) with an accuracy of 78.42\% \cite{moghaddasi2022novel}.

Related work has been done by Riccio et al.\ \cite{riccio2022atrial}, who developed ``eigenvalue dominance ratio'' maps, which are based on the singular values of the data matrix (or equivalently the eigenvalues of the associated sample covariance matrix) constructed from unipolar (catheter) electrograms. 
This data matrix contains the time-domain traces of each electrode. The method requires to time-align these traces by estimating the local activation time in each trace, which is done via an iterative process that maximizes the cross-correlations of the traces. After proper time alignment, the method detects the similarity of the AP morphologies, with the goal of the detection of fibrotic areas. The main distinction with our work is ($i$) it requires time-alignment, which is not always easy to achieve and relies on an underlying parametric model that does not account for fractionation; ($ii$) it works in the time domain rather than on the amplitude spectra, hence includes more phase information. This makes the methods not directly comparable.

In our proposed method, we took the element-wise absolute values of the Fourier spectra. This removes time-delay effects and allowed us to study changes in the morphology's dispersion without estimating the LAT. At the same time, the LAT (or phase of the $\bD$-matrix) can be regarded as independent, complementary information. It thus makes sense to look at both the activation map and the $\sigma_2$ map together. 

\subsection{Limitations and future work}
The proposed method computes the Fourier spectra of the measured signals and takes the absolute value. This step suppresses half the information present in these signals. In particular, the suppressed phase has all information on the local activation times. Thus, the proposed $\sigma_2$-map has independent information from the traditional activation maps. Future research should address the integration of these two feature maps, and relate them to the hidden electropathological parameters of the tissue such as conductivity.

Our study has shown that there are clear differences in normalized $\sigma_2$ between SR and AF-type array measurements. However, the results in Fig.\ \ref{boxplot} show that it is hard to propose a fixed threshold to distinguish between these two cases. Such a threshold would vary between 0.2 and 0.25 depending on location and other factors, such as the height of the electrodes above the tissue ($z_0$). Similarly, the BSP data (Fig.\ \ref{BSP_L1_L2}) has shown that at some locations, no difference is found. Thus, array placement is an issue that needs further study. 

We have shown $\sigma_2$-maps based on $3\times 3$ electrodes and a single beat. These highlight locations that deserve further attention. Alternatively, we have also shown plots (Fig.\ \ref{s_val_9loc} and Fig.\ \ref{EGM_BSP} ) where the normalized $\sigma_2$ of the entire array is computed (in a sense, averaging over space), and we could average that over multiple beats. That allows to compress a larger data set into a single feature. An open question is whether this averaging will dilute the differences between SR and AF such that this feature is not sufficiently discriminative anymore.

For the $\sigma_2$-maps such as Fig.\ \ref{s2_sr_clinical} and \ref{s2_af_clinical}, we have shown that areas of increased $\sigma_2$ correspond to EGMs with double potentials or related irregularities. We did not demonstrate the reverse, namely that {\em all} irrregular EGMs are highlighed in the $\sigma_2$-map. 

\section{Conclusion}\label{sec5}
In this paper, we developed a method for analyzing EGM and multi-lead ECG data. The method is non-parametric and requires little preprocessing. We have shown that the singular values of the processed data matrix give information on inhomogeneity of the AP morphologies, and the related $\sigma_2$-map points at areas subject to fractionation and block. The method gives a clear distinction between heartbeats in SR and AF. 

Further, experiments using simultaneous EGM and multi-lead ECG measurements showed that the singular values of the heartbeats during AF are higher than during SR, and that this difference is more pronounced for the ECG data than for the EGM data, if the electrodes are positioned at favorable locations.

Our related results show that the proposed singular value features can be a useful indicator to evaluate AF.

\section{Acknowledgements}
This research was funded in part by the Medical Delta Cardiac Arrhythmia Lab (CAL), The Netherlands. The authors would like to thank Dr.\ Frans B.S. Oei, cardio-thoracic surgeon at the Erasmus Medical Center (EMC), for his valuable suggestions on the cardio-thoracic measurements.

\appendix
\subsection{Proof of the claim in Section \ref{sec:flat}}
   We prove this for a continuous cell distribution in 1D space. For a traveling plane wave, the cell voltage as function of position $x$ is
\[
   c(x,t) = s(t-\frac{x}{v})
\]
   where $v$ is the propagation velocity. (Equivalently, the propagation delays $\tau$ are a linear function of position $x$.) For simplicity, we assume cells have an equal gain normalized to 1. In the frequency domain this then becomes
\[
   \tilde{c}(x,\omega) = S(\omega) e^{-j \omega x/v}.
\]
   The spatial response of an electrode centered at location 0 is some function $f(x)$; the example in (\ref{egm}) would in this context be $f(x) = a/|x|$. Assuming the response is linear space-invariant, the electrode voltage measured at a location $y$ is the convolution integral
\[
    m(y,t) = \int c(x,t) f(y-x) dx = \int c(y-x,t) f(x) dx
\]
   In frequency domain, this becomes
\begin{align*}
   \tilde{m}(y,\omega) &= \int S(\omega) e^{-j \omega (y-x)/v} f(x) dx
   \\
   &= S(\omega) e^{-j\omega y/v} \int e^{j\omega x/v} f(x) dx
   \\
   &=: S(\omega) e^{-j\omega y/v} I(\omega) 
\end{align*}
   Taking the absolute value gives
\[
  |\tilde{m}(y,\omega)| = |S(\omega) I(\omega)|
\]
    which is not a function of position $y$ anymore. Sampling over $y$ and $\omega$, the corresponding measurement matrix $\bD$ will be rank 1. Generally, $\bD$ will be rank 1 if we are able to factorize $|\tilde{m}(y,\omega)|$ as $|\tilde{m}(y,\omega)|=A(y)B(\omega)$. This also shows that we can permit electrodes to have different gains (as long as they are frequency-independent): this results in a different factor $A(y)$ but will not increase the rank.

    Generalizing the derivation, we observe that if cells have unequal gains $a(x)$, i.e., $c(x,t) = a(x) s(t-\frac{x}{v})$, we will have rank 1 only if we can factorize $a(y-x)$ into separate factors depending on $y$ and $x$. It follows that $a(x)$ must be of the form $a(x) = e^{\alpha x}$. For small $\alpha$, this can be linearized to $a(x) = 1 + \alpha x$. Thus, we can permit a small gain gradient over the cells. 

    If the wavefront is curved, the delays are a nonlinear function of $x$, e.g. $\tau = \left(x + \frac{a}{x} \right)/v$. It is easily seen that in this case, $|\tilde{m}(y,\omega)|$ does not factorize, so that $\bD$ will have rank higher than 1.

\bibliographystyle{IEEEtran}

\end{document}